\documentclass[a4paper]{cas-sc}

\usepackage[authoryear]{natbib}

\def\tdc#1{\csdef{#1}{\textdc{\lowercase{#1}}\xspace}}
\tdc{WGM}
\tdc{QE}

\usepackage{caption}
\usepackage{booktabs}
\usepackage{algorithm}  
\usepackage{algorithmic}
\usepackage{longtable} 

\makeatletter
\newenvironment{breakablealgorithm}
{
		\begin{center}
			\refstepcounter{algorithm}
			\hrule height.8pt depth0pt \kern2pt
			\renewcommand{\caption}[2][\relax]{
				{\raggedright\textbf{\ALG@name~\thealgorithm} ##2\par}%
				\ifx\relax##1\relax 
				\addcontentsline{loa}{algorithm}{\protect\numberline{\thealgorithm}##2}%
				\else 
				\addcontentsline{loa}{algorithm}{\protect\numberline{\thealgorithm}##1}%
				\fi
				\kern2pt\hrule\kern2pt
			}
		}{
		\kern2pt\hrule\relax
	\end{center}
}
\makeatother
\makeatletter
\newcommand{\figcaption}{\def\@captype{figure}\caption}
\newcommand{\tabcaption}{\def\@captype{table}\caption}
\makeatother

\begin{document}
\let\WriteBookmarks\relax
\def\floatpagepagefraction{1}
\def\textpagefraction{.001}
\let\printorcid\relax

\shorttitle{QTypeMix: Enhancing Multi-Agent Cooperative Strategies through Heterogeneous and Homogeneous Value Decomposition}    

\shortauthors{Songchen Fu et al.}  

\title [mode = title]{QTypeMix: Enhancing Multi-Agent Cooperative Strategies through Heterogeneous and Homogeneous Value Decomposition}  

%

\author[author1,author2]{Songchen Fu}
\ead{fusongchen@hccl.ioa.ac.cn}
\credit{Conceptualization, Data curation, Formal analysis, Methodology, Project administration, Investigation, Software, Visualization, Writing – original draft, Writing – review and editing}

\author[author1,author2]{Shaojing Zhao}
\ead{zhaoshaojing@hccl.ioa.ac.cn}
\credit{Methodology, Investigation, Writing – review and editing}

\author[author1,author2]{Ta Li}
\cormark[1]
\ead{lita@hccl.ioa.ac.cn}
\credit{Supervision, Methodology, Writing – review and editing}

\author[author1,author2]{Yonghong Yan}
\ead{yanyonghong@hccl.ioa.ac.cn}
\credit{Supervision, Conceptualization, Resources}

\affiliation[author1]{organization={Laboratory of Speech and Intelligent Information Processing, Institute of Acoustics},
            city={Beijing},
            country={China}}
\affiliation[author2]{organization={University of Chinese Academy of Sciences},
            city={Beijing},
            country={China}}

\cortext[cor1]{Corresponding author}



\begin{abstract}
In multi-agent cooperative tasks, the presence of heterogeneous agents is familiar. Compared to cooperation among homogeneous agents, collaboration requires considering the best-suited sub-tasks for each agent. However, the operation of multi-agent systems often involves a large amount of complex interaction information, making it more challenging to learn heterogeneous strategies. Related multi-agent reinforcement learning methods sometimes use grouping mechanisms to form smaller cooperative groups or leverage prior domain knowledge to learn strategies for different roles. In contrast, agents should learn deeper role features without relying on additional information. Therefore, we propose QTypeMix, which divides the value decomposition process into homogeneous and heterogeneous stages. QTypeMix learns to extract type features from local historical observations through the TE loss. In addition, we introduce advanced network structures containing attention mechanisms and hypernets to enhance the representation capability and achieve the value decomposition process. The results of testing the proposed method on 14 maps from SMAC and SMACv2 show that QTypeMix achieves state-of-the-art performance in tasks of varying difficulty. 
\end{abstract}




\maketitle

\section{Introduction}
\label{1}

As one of the essential branches in multi-agent reinforcement learning (MARL), cooperative MARL has become a solution for many unmanned robot cluster tasks \cite{wang2022task, xu2018multi, wang2020multi}. Unlike competitive and hybrid tasks, collaborative tasks guide the actions of each agent through team rewards to ensure that the goals of each agent are consistent and that there is no competitive relationship. In the distributed training with decentralized execution (DTDE) paradigm, each agent learns independently without utilizing explicit information exchange or global observation \cite{lauer2000algorithm, de2020independent}. However, this approach of treating other agents as part of the environment is non-stationary \cite{zhang2021multi, wong2023deep}. The centralized training with centralized execution (CTCE) paradigm allows for real-time information exchange or global observation to learn the joint policy of all agents through centralized actuators \cite{gupta2017cooperative}. Although the CTCE paradigm alleviates the non-stationary problem, most multi-agent systems have partial observability and communication constraints, making it difficult to achieve good adaptation \cite{gronauer2022multi}. Therefore, the centralized training with distributed execution (CTDE) paradigm \cite{kraemer2016multi} has been widely applied, which combines the advantages of the previous two paradigms and is easy to apply to various tasks \cite{oliehoek2008optimal, foerster2016learning, sunehag2017value, foerster2018counterfactual}. \par

Similar to single-agent reinforcement learning (RL), there are two main types of MARL algorithms in the CTDE paradigm: value-based and policy-based methods. The value-based methods mainly study how to decompose the joint value function learned from the team reward function into local value functions, forming individual strategies for each agent. This type of method is generally referred to as Value Function Factorization (VFF) methods, such as VDN \cite{sunehag2017value} and QMIX \cite{rashid2020monotonic}, which have shown good performance in many intelligent agent tasks but are challenging to apply to tasks with continuous action spaces. On the other hand, under the CTDE paradigm, policy-based methods directly learn parameterized policies and mainly adopt a centralized-cirtic-distributed-actor structure \cite{foerster2018counterfactual, lowe2017multi, yu2022surprising}. Although policy-based methods can be applied to tasks in continuous action spaces, they can easily lead to local optima under the guidance of team rewards. \par

This article mainly researches value-based methods under the CTDE paradigm. In recent years, many VFF methods have been proposed, among which the two mainstream types of ideas are: increasing the representation ability of VFF by continuously reducing constraints \cite{sunehag2017value, rashid2020monotonic, son2019qtran, yang2020qatten, hu2021rethinking,shen2022resq} and improving the algorithm's exploration ability for policy space and environment interaction \cite{mahajan2019maven, lyu2020likelihood, sun2021dfac, pmlr-v139-gupta21a}. The purpose of increasing VFF representation ability is to find more accurate credit assignment methods, and the purpose of growing exploration ability is to improve the sampling efficiency of agents. However, the primitive flat VFF scheme \cite{phan2021vast} has two main issues. First, it needs to work on accurately learning the impact of each agent's actions on the team value function through a global state. Second, it faces challenges distinguishing agents' contributions with varying abilities in heterogeneous agent scenarios. These issues will significantly decrease performance as the number of agents in the environment increases. \par

In earlier years, research in natural systems has demonstrated the effectiveness of grouping and division of labor in collaborative labor \cite{gordon1996organization, jeanson2005emergence, wittemyer2007hierarchical}. Inspired by this, methods of grouping according to specific prior rules or adaptive grouping have also been introduced into MARL, which makes individual contributions to the team more accessible to learn. Specifically, some methods group agents using prior knowledge \cite{lhaksmana2018role, jiang2021multi}, some methods automatically group agents based on specific rules during interactions with the environment \cite{shao2022self, zang2024automatic}, and other methods learn the individuality or roles of the agents \cite{wang2020roma, wang2020rode, jiang2021emergence}. However, due to the ample joint state space of multi-agent systems and the high exploration randomness, learning how to group agents entirely through neural networks places high demands on the design of the loss function. Therefore, we adopt a novel approach that has yet to be explored in prior studies. Rather than grouping agents based on functionality, proximity, or alignment with short-term goals, we focus on defining the roles various agents should fulfill within the team. \par

In this paper, we investigate how to make the learning process from individual action-value functions to the joint action-value function more directional from a new perspective, thereby obtaining a more accurate representation of the individual's impact on the team. To address the abovementioned issues, we propose a novel type-related VFF method, QTypeMix, which performs hierarchical value decomposition based solely on agent types provided by the environment or human input. Specifically, QTypeMix uses the global state to decompose the joint action-value function into type-level value functions and then uses the global state, local observations, and other information to decompose the type-level value functions into local utilities. Additionally, we design a feature extractor that extracts type-related observation embedding from each agent's historical observations to generate the network weights for the second-layer value decomposition process. It is important to note that while our method may seem similar to group-based VFF methods, it differs fundamentally. Rather than dividing agents into teams by type, we guide each to learn strategies most valuable for their specific type in the given task. Our contributions are as follows: \par

(1) We propose a novel dual-layer value decomposition method, QTypeMix. This method innovatively leverages local observations and type-related embeddings to provide more direct guidance for the value function decomposition process. Compared to existing methods that only use global states or no information, QTypeMix exploits the common features in the historical observations of agents of the same type. This approach guides the network to different functionalities, increasing learning efficiency. \par

(2) We designed a new feature extractor to derive type-related embeddings based on each agent's historical observations, which are trained through an additional loss function. The type-related embeddings guide the neural network in decomposing type-level value functions into local utilities, providing additional information for policy optimization. \par

(3) We employed various advanced network architectures to maximize the neural network's representational capacity and integrated the algorithm into mainstream MARL frameworks to ensure its generalization and reproducibility. \par

The experiments in this paper are based on selected scenarios from the StarCraft Multi-Agent Challenge (SAMC) \cite{samvelyan2019starcraft} and SAMCv2 \cite{ellis2024smacv2}. The final results show that QTypeMix matches or exceeds the performance of existing SOTA methods in most scenarios and demonstrates outstanding performance in scenarios with many agent types. Our code is available at \href{https://github.com/linkjoker1006/pymarl3}{https://github.com/linkjoker1006/pymarl3}. \par

\section{Related Work}
\label{2}

\textbf{Centralized Training with Distributed Execution:} Due to the non-stationarity and partial observability problems in multi-agent systems, MARL requires the design of an efficient multi-agent training scheme. The CTDE paradigm cleverly combines the advantages of the DTDE and CTCE paradigms, allowing agents to access global or additional information during training while only relying on local observations to take actions during execution. To address the issue in the DTDE paradigm, where agents struggle to distinguish between environmental changes and changes caused by other agents, CTDE mitigates non-stationarity and partial observability by allowing access to global states or other agents' observations. Conversely, to tackle the issue in the CTCE paradigm where agents must obtain global information during execution, CTDE uses decentralized execution, significantly increasing the algorithm's flexibility and reducing environmental requirements. Additionally, the CTDE paradigm supports both value-based \cite{sunehag2017value, rashid2020monotonic, son2019qtran, yang2020qatten, shen2022resq, jianye2022boosting, mahajan2019maven, sun2021dfac, pmlr-v139-gupta21a} and policy-based \cite{foerster2018counterfactual, lowe2017multi, yu2022surprising, mguni2021ligs, chen2023ljir} methods. The method proposed in this paper is a value-based MARL method under the CTDE paradigm. \par

\textbf{Value Function Factorization:} The VFF method factorizes the joint action-value function into local utilities through mixing networks, making it the most common type of value-based MARL method. VDN uses a linear decomposition method to decompose the joint action-value function. QMIX generates a non-linear decomposition scheme under monotonicity constraints, enhancing the representation capability of the mixing network through a hypernetwork. QTRAN \cite{son2019qtran} abandons the monotonicity and additivity assumptions, performing value decomposition by estimating error terms and proposing the Individual-Global-Max (IGM) principle. Qatten \cite{yang2020qatten} derives the general form of global and individual Q-values using Taylor expansion, freeing itself from assumptions about their relationship. HPN-QMIX \cite{jianye2022boosting} introduces permutation invariance (PI) and permutation equivariance (PE) into QMIX, significantly improving model representation capability through a hypernetwork and effectively reducing the state space of MARL. With comprehensive theoretical derivation and ingenious algorithm implementation, this method achieves substantial performance improvements over baseline methods in the SMAC and SMACv2 environments. Besides relaxing constraints and enhancing the representation capability of the mixing network, another mainstream approach is to increase agent exploration ability or improve data collection efficiency. MAVEN \cite{mahajan2019maven} addresses the inefficiency in exploration caused by QMIX constraints through committed exploration by introducing a latent space for hierarchical control. This method mixes value-based and policy-based approaches, conditioning the value-based policy on latent variables controlled by a hierarchical policy. \par

\textbf{Group-based and Role-based MARL:} Group-based or role-based MARL methods utilize predefined or learned group and role information to guide value decomposition or policy optimization, often implemented through multi-level structures. The type-based hierarchical group communication (THGC) \cite{jiang2021multi} utilizes prior or predefined rules in fields such as location, functionality, and health to the group and maintains the group’s cognitive consistency through knowledge sharing. The Self Organized Group (SOG) \cite{shao2022self} is featured with a conductor election (CE) and a message summary (MS) mechanism, which enable the algorithm to have zero-shot generalization ability in terms of the dynamic number of agents and the varying partial observability. The Group oriented Multi Agent Reinforcement Learning (GoMARL) \cite{zang2024automatic} enables models to automatically learn how to group without domain knowledge guidance, decompose the joint action-value function into group-wise value functions, and further guide agents to improve their policies in a fine-grained fashion. The Emergence of Individuality (EOI) \cite{jiang2021emergence} trains a probabilistic classifier that predicts the probability distribution of agents given observations using intrinsic rewards. Agents tend to visit observations they are familiar with, promoting the emergence of individuality. RODE \cite{wang2020rode} proposed a role-based dual-level structure where a role selector periodically searches the role space, and role policies learn within the decomposed joint action space. Unlike existing methods, QTypeMix approaches from the perspective of agent types, extracting unique observational features for each type to guide the value decomposition process. The algorithm only utilizes the prior knowledge of agent types, which is undoubtedly easily provided by any environment supporting heterogeneous agents. \par

\section{Background}
\label{3}

\textbf{Decentralized Partially Observable Markov Decision Process:} This paper focuses on fully cooperative multi-agent tasks, where all agents in the environment attempt to maximize a joint reward function while having different individual goals \cite{wong2023deep}. Similar to modeling single-agent dynamic decision-making in stochastic environments using partially observable Markov decision process (POMDP), fully cooperative multi-agent tasks are typically modeled as decentralized partially observable Markov decision process (Dec-POMDP) \cite{bernstein2002complexity, oliehoek2016concise}. Dec-POMDP is defined as a tuple $G = \langle S, U, P, r, Z, O, n, \gamma \rangle$. $s \in S$ represents the global state of the environment. Each agent $a_i \in \mathcal{A} = \{a_1, \ldots, a_n\}$ takes an action $u^i_t \in U$ at timestep $t$, forming the joint action $u_t \in U^n$. The state transition distribution $P(s_{t+1} | s_t, \boldsymbol{u}_t): S \times U^n \times S \rightarrow [0, 1]$ governs the environment's state transitions caused by the joint actions. $r(s, \boldsymbol{u}): S \times U^n \rightarrow \mathbb{R}$ defines the team reward function shared by all agents and $\gamma \in [0, 1)$ is the discount factor. In a partially observable setting, agent $a_i$ cannot access the global state and can only sample local observation $z^i \in Z$ through the observation function $O(s,i): S \times A \rightarrow Z$. Therefore, the action-observation history of agent $a_i$ is $\tau^i \in T \equiv (Z \times U)^*$ on which it conditions a policy $\pi^i(u^i | \tau^i): T \times U \rightarrow [0,1]$. The joint policy $\pi $ is based on the joint action-value function $Q^\pi(s_t | \boldsymbol{u}_t) = \mathbb{E}_{s_{t+1}:\infty, \boldsymbol{u}_{t+1}:\infty}[R_t | s_t, \boldsymbol{u}_t]$, where $R_t = \sum_{k=0}^{\infty}\gamma^k r_{t+k}$ is the discounted reward. \par

\textbf{QMIX:} In RL, Deep Q-Network (DQN) \cite{mnih2015human} is the most typical value-based method. This method leverages deep neural networks to approximate the optimal action-value function $Q^*(s,a) = \max_{\pi} \mathbb{E} [ r_t + \gamma r_{t+1} + \gamma^2 r_{t+2} + \cdots \mid s_t = s, a_t = a, \pi ]$. During training, DQN updates the policy's Q-value function by sampling transitions $(s, u, r, s')$ with batch size $B$ from the replay buffer $D$ and minimizing the squared temporal difference (TD) error:

\begin{equation}\label{eq1}
\mathcal{L}(\theta) = \sum_b^{B} \left[ \left( y_b^{DQN} - Q(s,u; \theta) \right)^2 \right],
\end{equation}
where $y^{DQN} = r + \gamma \max_{u'} Q(s',u'; \theta^-)$. $\theta^-$ are the parameters of the target network that are periodically copied from $\theta$ and kept constant for several iterations. \par

After applying DQN to multi-agent systems, each agent has its independent action-value function $Q_i$. Although allowing multiple agents to update their policies during training independently can be effective in terms of results \cite{tampuu2017multiagent}, the non-stationarity caused by the impact of other agent's actions in the environment makes convergence unguaranteed. QMIX \cite{rashid2020monotonic} is one of the most classic value decomposition algorithms, and the algorithm proposed in our paper is also an improvement based on this framework. It employs a mixing network to estimate the joint action-value $Q_{tot}$ as a monotonic combination of individual Q-value $Q_i$ of each agent. By controlling the non-negativity of the weights, QMIX maintains the consistency between centralized and distributed policies. Therefore, its monotonicity constraint can be expressed as:

\begin{equation}\label{eq2}
    \frac{\partial Q_{tot}}{\partial Q_a} \geq 0, \quad \forall a.
\end{equation}\par

This monotonicity constraint enables QMIX to meet the important Individual-Global-Max (IGM) principle proposed in subsequent work \cite{son2019qtran}, which states that for a joint action-value function $ Q^*(\boldsymbol{\tau}, \boldsymbol{u}): T^n \times U^n \rightarrow \mathbb{R}$, if there exist individual action-value functions $[Q_i(\tau, u): T \times U \rightarrow \mathbb{R}]^n_{i=1}$, the following holds:

\begin{equation}\label{eq3}
\arg \max_{\boldsymbol{u}} Q^*(\boldsymbol{\tau}, \boldsymbol{u}) = 
\begin{pmatrix}
\arg \max_{u^1} Q_1(\tau^1, u^1) \\
\vdots \\
\arg \max_{u^n} Q_n(\tau^n, u^n)
\end{pmatrix},
\quad \forall \boldsymbol{\tau}, \boldsymbol{u}.
\end{equation}\par

QMIX's mixing network is also implemented through a four-layer structure similar to the hypernetworks \cite{ha2016hypernetworks}. It uses the global state to generate the weights and biases for the value decomposition process, allowing QMIX to access global information during training. \par

\textbf{Attention mechanism:} Since its introduction, the attention mechanism \cite{vaswani2017attention} has been widely applied in various research fields, and many MARL works \cite{iqbal2019actor, das2019tarmac, yang2020qatten, zhang2022multi, pu2022attention} have also utilized this concept. Qatten utilizes a multi-head attention mechanism to replace the mixing network in QMIX, approximating $Q_{tot}$. AERL \cite{pu2022attention} employs the graph attention operator to handle complex agent interactions and capture temporal correlations. An attention function can be described as mapping a query and a set of key-value pairs to an output, where the query, keys, values, and output are all vectors. The output is computed as a weighted sum of the values, where a compatibility function of the query with the corresponding key computes the weight assigned to each value. In practice, we compute the attention function on a set of queries simultaneously, packed into a matrix $Q$. The keys and values are also loaded into matrices $K$ and $V$. We compute the matrix of outputs as:
\begin{equation}\label{eq4}
\text{Attention}(Q, K, V) = \text{softmax}\left(\frac{QK^T}{\sqrt{d_k}}\right)V.
\end{equation} 
where $\sqrt{d_k}$ represents the dimension of the values. This paper employs the multi-head attention mechanism to jointly enable the model to attend to information from different representation subspaces. \par

\textbf{Hyper Policy Network:} To overcome the curse of dimensionality in MARL, where the state space grows exponentially with the number of agents, \cite{jianye2022boosting} proposed Hyper Policy Network (HPN). This method uses inductive biases of permutation invariance (PI) and permutation equivariance (PE) to reduce the multi-agent state space, significantly improving existing MARL methods' performance and learning efficiency. Regarding results, the HPN series algorithms can be considered to have achieved state-of-the-art (SOTA) performance on both SMAC and SMACv2. Therefore, we use this class of methods as the primary baseline. In recognition of the outstanding contribution of HPN, the algorithm proposed in this paper adopts the same network structure for the module that extracts each agent's observation features. \par

\section{Methodology}
\label{4}

In this section, we introduce the overall architecture of QTypeMix and provide a detailed description of its implementation. In QTypeMix, types are defined as labels that distinguish agents with different capabilities or attributes, meaning that agents of the same type can be considered identical. For instance, in the SMAC environment where agent type information is provided, we use this information to categorize different unit types. Agents of the same unit type (e.g., marines) are considered completely identical. In environments where agent type information is not provided (e.g., Multi-Agent Particle Environment (MPE) \cite{lowe2017multi}), we categorize agents based on attribute values such as size, acceleration, and maximum speed. If a multi-agent system has $n$ agents $\mathcal{A} = \{a_1, \ldots, a_n\}$, we can divide them into $m$ types $\mathcal{A} = \{\mathcal{T}_1, \ldots, \mathcal{T}_m\}, 0 < m < n$, where $\mathcal{T}_k = \{a_{k_1}, \ldots, a_{k_{n_k}}\}, k \in \{1,2,\ldots,m\}, n_k = len(\mathcal{T}_k)$. After categorizing by type, each agent belongs exclusively to one type, which means for $\forall k, l \in \{1,2,\ldots,m\}$ and $k \neq l$, $\mathcal{T}_k \cap \mathcal{T}_l = \emptyset$. Although the experiments in this paper are conducted in SMAC and SMACv2 environments, the above definition of types allows QTypeMix to be easily applied to most multi-agent simulation environments. \par

\subsection{Simple type-wise value decomposition}
\label{4.1}

\begin{figure*}[htbp]
	\centering
		\includegraphics[scale=0.6]{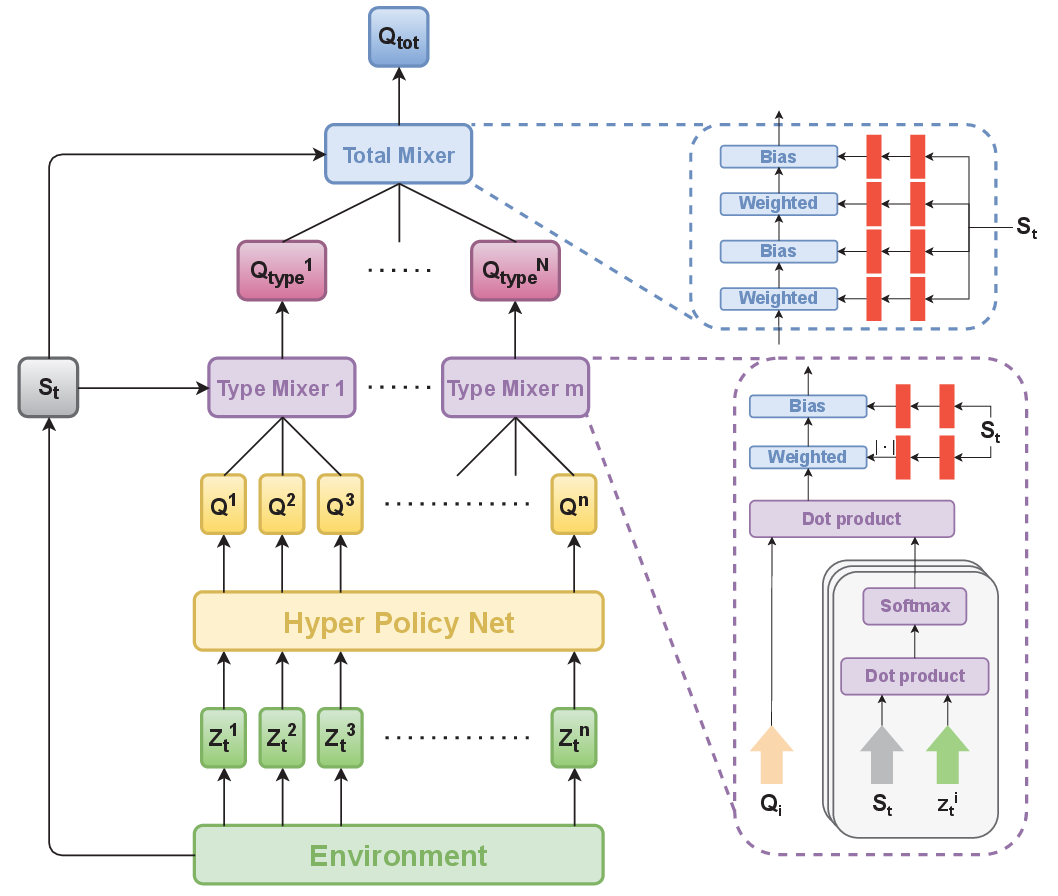}
	  \caption{Diagram of our straightforward type-wise value decomposition method (QTypeMix-B).}\label{fig1}
\end{figure*}

A straightforward way to allow the model to distinguish between agents of various types is to introduce an additional layer into the flat VFF scheme.  Using QMIX as an example, the algorithm represents the joint action-value function as a weighted sum of individual utilities plus a bias. Based on this setup, we add an extra value decomposition step according to the types of agents. The entire value decomposition process becomes $Q_{tot}$, first decomposed into $Q_{type}^k$, then decomposed into individual utilities $Q^i$. As shown in Fig.~\ref{fig1}, this foundational algorithm is referred to as QTypeMix-B(hereafter referred to as QTypeMix-B). By introducing the Type Mixer, we aim for the model to adopt similar value decomposition standards for agents of the same type and ultimately emphasize the differences in value contributed by different types of agents to the task. Like the Total Mixer, the Type Mixer also requires additional global state information from the environment during training, which is unnecessary during execution. \par

\begin{figure*}[htbp]
	\centering
		\includegraphics[scale=0.6]{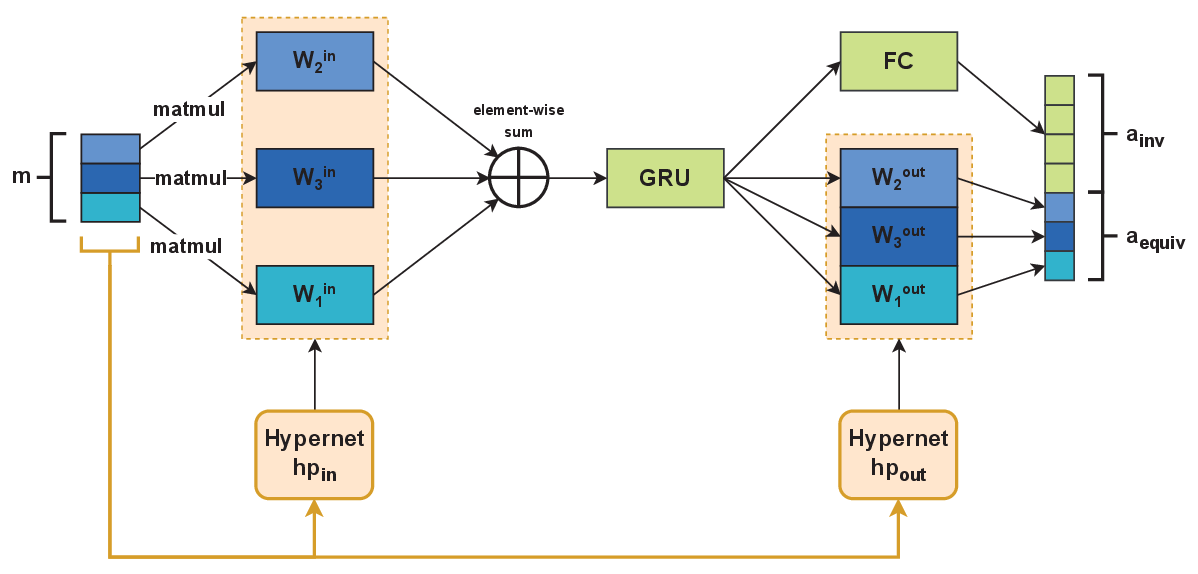}
	  \caption{Diagram of hyper policy network.}\label{fig2}
\end{figure*}

As shown in Fig.~\ref{fig1}, during training, the environment provides the global state and each agent's local observations $Z_t^i$ at each time step. The local observations $Z_t^i$ are first processed through an HPN to compute individual utilities, with the specific implementation detailed in Fig.~\ref{fig2}. The HPN primarily consists of two hypernetworks, $hp_{in}$ and $hp_{out}$. $hp_{in}$ generates the weights for a linear layer, sums the results, and then passes them through a GRU layer followed by a linear layer, achieving PI for the output actions. $hp_{out}$ multiplies the GRU results with network parameters generated based on the input and then adds a bias, resulting in PE output actions. In SMAC and SMACv2, the actions related to the agents' movements should satisfy PI, while those about attacking enemies should satisfy PE. For more details about HPN, please refer to \cite{jianye2022boosting}. In this paper, we use HPN merely as a tool. \par

After obtaining the individual action-values of agents at the current time step, $m$ Type Mixers process the utilities of agents belonging to their respective types. Finally, the type-specific Q-values output by the Type Mixers are fed into the Total Mixer to obtain the joint action-value $Q_{tot}$. We implement the Type Mixer using a multi-head attention mechanism cascaded with a hypernetwork guided by the global state. The Total Mixer, on the other hand, is entirely implemented using hypernetworks. Comparing the implementations of the two Mixers, we can see that introducing the attention mechanism allows the model to better focus on each agent's observations. Since each attention mechanism receives inputs from agents of the same type, we expect the model to learn the observation characteristics of each type of agent, thereby influencing the type-specific Q-values. Upon examining the two value decomposition stages of QTypeMix-B, it can be observed that each Type Mixer decomposes the type-specific action-value of homogeneous agents. In contrast, the Total Mixer decomposes the joint action-value into type-specific action-values. Therefore, we refer to the former as the homogeneous stage of value decomposition and the latter as the heterogeneous stage. With this, we have completed the implementation of QTypeMix-B. The relevant experimental results are presented in Section~\ref{5}. \par

\subsection{Additional type information extraction}
\label{4.2}
\begin{figure*}[htbp]
	\centering
		\includegraphics[scale=0.6]{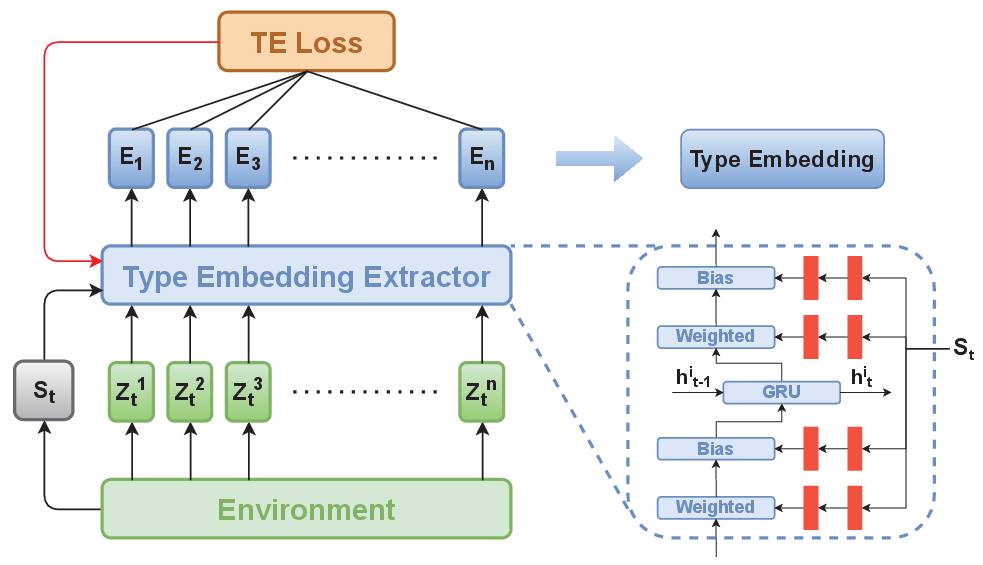}
	  \caption{Diagram of additional type information extraction.}\label{fig3}
\end{figure*}

In Section~\ref{4.1}, we proposed an intuitive and straightforward type-wise value decomposition method, aiming to better evaluate the action-values of different types of agents through the Type Mixer. In the attention mechanism of the Type Mixer, we use the agent's current local observation as the key. However, in multi-agent cooperative environments, relying solely on the local observation at the current time step makes it difficult to find associations with the type. For example, observing their local states at a single time step in an environment with two agents with different running speeds does not allow for distinguishing between them. If we can access the historical observations of each agent, we can quickly identify which agent runs faster and then allocate tasks more appropriately based on the mission objectives. \par

As shown in Fig.~\ref{fig3}, after obtaining the global state at time step $t$ and the local observations of each agent from the environment, we extract the type embedding $E_t^i$ for each agent from their historical observations $\{Z_0^i, Z_1^i, \ldots, Z_t^i\}$ using the Type Embedding Extractor. It consists of GRU units and linear layers generated by hypernetworks. To guide the network in updating in the desired direction, we design the TE loss in addition to the standard RL training process, which means that the updates for TE loss and TD Loss are mutually independent. The specific definition of the TE loss is as follows:
\begin{equation}\label{eq5}
\begin{aligned}
\mathcal{L}_{TE} (\theta_{te}) &= \sum_b^{B} \left( \sum_i^n \sum_j^n I(i, j) \cdot \cos \left( E_i(Z^i, S; \theta_{te}), E_j(Z^j, S; \theta_{te}) \right) \right), \\
I(i, j) &= 
\begin{cases}
-1, & \text{if } a_i \in \mathcal{T}_k, a_j \in \mathcal{T}_k. \\
1, & \text{if } a_i \in \mathcal{T}_k, a_j \notin \mathcal{T}_k.
\end{cases}
\end{aligned}
\end{equation}
We calculate the cosine similarity between the type embeddings of every two agents and use an indicator function $I$ to control the positive or negative contributions, aiming to maximize the embedding differences between different types and minimize the embedding differences within the same type. Under the guidance of the TE loss, the type embeddings will incorporate more information related to the agents' types. For example, type A agents may have higher mobility, type B agents may possess greater attack power, and type C agents may have higher health points. \par

\subsection{Algorithm overview}
\label{4.3}
\begin{figure*}[htbp]
	\centering
		\includegraphics[scale=0.6]{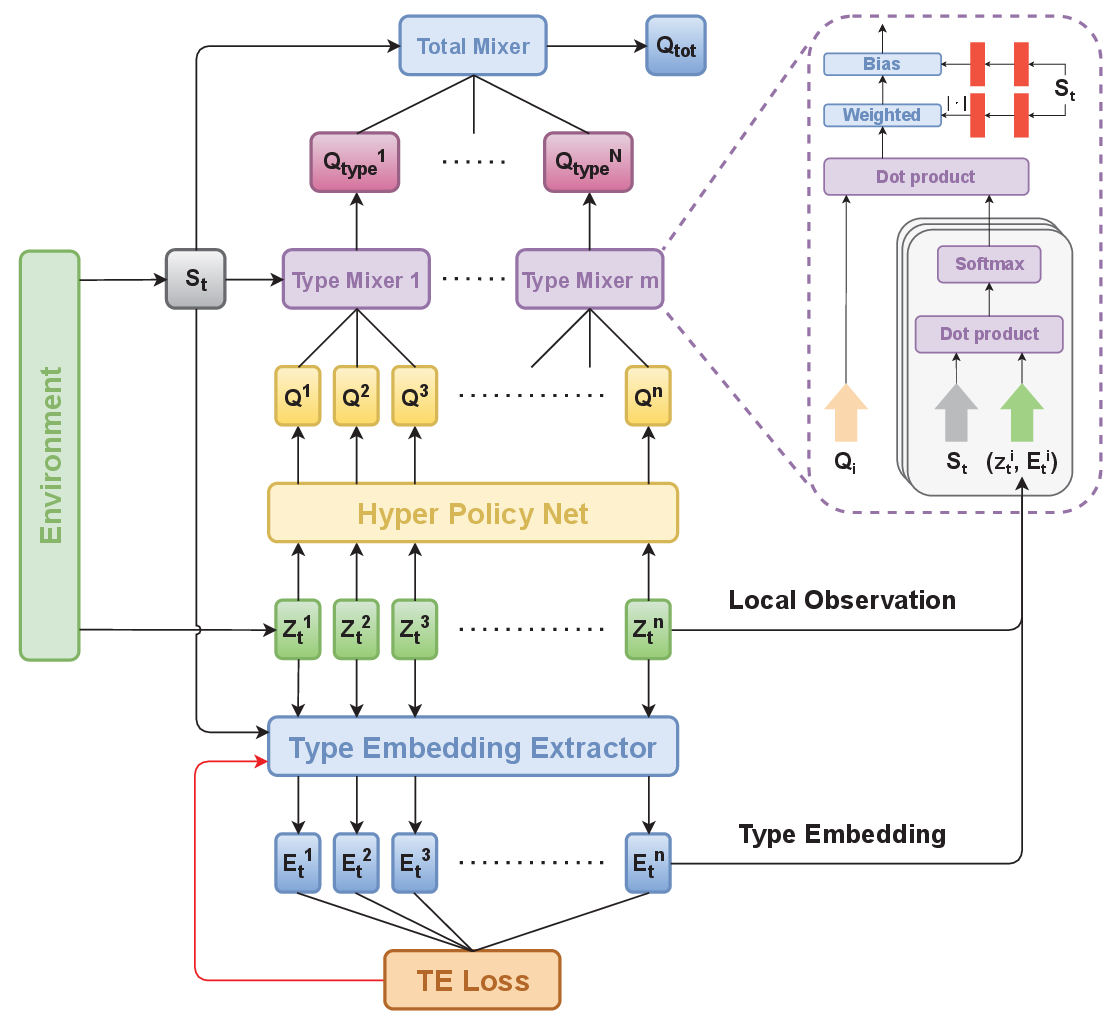}
	  \caption{Diagram of QTypeMix.}\label{fig4}
\end{figure*}

After introducing QTypeMix-B and the Type Embedding Extractor, we will present the overall architecture of QTypeMix. As shown in Figure 4, the entire algorithm includes two mixing networks and an embedding extractor, all of which involve using hypernetworks to generate weights and biases. To prevent the weights generated by the hypernetworks from being zeroed out by the activation function when they are negative, QTypeMix, like QMIX, uses ELU as the non-linear function instead of ReLU. Regarding the workflow, the agents' local observations $Z_t^i$ first pass through the HPN and the Type Embedding Extractor to obtain local utilities $Q_t^i$ and type embeddings $E_t^i$. In the homogeneous value decomposition stage, each agent's local observation and type embedding are concatenated to form the key (K). The global state serves as the value (V), and the local utilities serve as the query (Q). These are processed through a multi-head attention mechanism to calculate the type utility. The weights and biases for the multi-head attention and the weights and biases for the weighted summation of the multi-head results are all generated by a hypernetwork that takes the global state $S_t$ as input. In the heterogeneous value decomposition stage, the multiple $Q_{type}^k$ are fed into the Total Mixer, where the weights and biases are entirely generated by the hypernetworks, to obtain the estimated joint action-value $Q_{tot}$. \par

As described earlier, the network update for QTypeMix consists of two backpropagation steps. Although QTypeMix has unique characteristics in the value decomposition process, the TD error used for updating the network is consistent with other value-based methods:
\begin{equation}\label{eq6}
\mathcal{L}_{TD}(\theta) = \sum_b^{B} \left[ \left( r + \gamma \max_{u'} Q_{tot}(s',u'; \theta^-) - Q_{tot}(s,u; \theta) \right)^2 \right].
\end{equation}
Since the gradient of the TD error passes through four neural networks, $\theta$ in Eq.~\ref{6} includes all the network parameters in QTypeMix: $(\theta_{te}, \theta_{hpn}, \theta_{m1}, \theta_{m2})$. That is when updating the TD error, we do not freeze the parameters of the Type Embedding Extractor. Instead, we add a weight $\alpha$ to allow the two loss functions to update together. We believe this approach can further ensure the direction of updating and learning efficiency of the Type Embedding Extractor. Therefore, the global optimization objective of QTypeMix is to minimize:
\begin{equation}\label{eq7}
\mathcal{L}(\theta) = \mathcal{L}_{TD}(\theta) + \alpha \cdot \mathcal{L}_{TE}(\theta_{te})
\end{equation}
The detailed process of the QTypeMix algorithm is outlined in Appendix~\ref{A}.

\section{Experiments}
\label{5}

\begin{table}[htbp]
\centering
\caption{Detailed test results of 6 algorithms on 14 maps.}\label{tbl1}
\begin{tabular}{cccccccc}
\toprule
Map & Difficulty & FT-VDN & FT-QMIX & HPN-VDN & HPN-QMIX & QTypeMix-B & QTypeMix \\
\midrule
2c vs. 64zg & Hard & 76.0(8.2) & 98.8(1.8) & 98.8(2.0) & 99.4(1.4) & 98.8(1.8) & \textbf{99.5(1.1)} \\
3s5z vs. 3s6z & Super Hard & 55.9(7.3) & 73.2(7.3) & 93.8(4.2) & 97.7(2.9) & 97.7(2.6) & \textbf{97.9(2.1)} \\
6h vs. 8z & Super Hard & 68.4(8.7) & 87.0(6.3) & 90.2(5.6) & 97.2(3.8) & 95.2(3.6) & \textbf{97.2(2.6)} \\
corridor & Super Hard & 81.8(8.5) & 92.8(4.3) & 97.4(2.6) & 97.3(2.0) & 97.0(3.0) & \textbf{97.7(2.9)} \\
MMM2 & Super Hard & 88.2(5.6) & 92.0(4.1) & 99.7(0.9) & 99.6(1.0) & 99.7(0.9) & \textbf{99.8(0.7)} \\
protoss 5 vs. 5 & SAMCv2 & 61.2(8.2) & 62.9(9.5) & 76.0(8.7) & 76.9(7.5) & 78.7(7.4) & \textbf{78.8(7.3)} \\
protoss 10 vs. 11 & SAMCv2 & 12.5(5.7) & 16.8(6.5) & 51.7(8.9) & 49.6(9.4) & 55.0(9.2) & \textbf{60.2(9.0)} \\
protoss 15 vs. 16 & SAMCv2 & 11.1(5.7) & 14.8(6.1) & 50.8(8.6) & 56.0(9.2) & 60.8(7.9) & \textbf{62.8(8.9)} \\
terran 5 vs. 5 & SAMCv2 & 65.6(8.3) & 73.4(8.3) & 74.6(8.7) & 75.6(7.9) & 76.8(8.2) & \textbf{77.3(7.4)} \\
terran 10 vs. 11 & SAMCv2 & 39.7(8.1) & 47.9(8.5) & 63.0(9.2) & 65.8(9.2) & 67.1(8.8) & \textbf{68.9(8.7)} \\
terran 15 vs. 16 & SAMCv2 & 39.3(9.1) & 50.7(9.4) & 58.5(8.5) & 65.4(9.7) & 67.1(8.8) & \textbf{72.7(8.5)} \\
zerg 5 vs. 5 & SAMCv2 & 54.6(8.3) & 60.3(8.7) & 65.8(8.5) & 68.3(7.4) & 70.5(7.6) & \textbf{71.6(7.3)} \\
zerg 10 vs. 11 & SAMCv2 & 42.3(9.0) & 43.2(9.5) & 58.5(9.6) & 51.1(8.6) & 58.0(9.7) & \textbf{61.5(8.9)} \\
zerg 15 vs. 16 & SAMCv2 & 41.5(9.3) & 50.5(9.3) & 59.1(8.1) & 58.8(8.5) & 58.8(9.4) & \textbf{60.9(8.8)} \\
\bottomrule
\end{tabular}
\end{table}

\textbf{Experimental Setup:} We select the commonly used SMAC \cite{samvelyan2019starcraft} and the more challenging SMACv2 \cite{ellis2024smacv2} as benchmarks to evaluate the performance of QTypeMix. According to the official recommendation, we use game version 4.6.2.69232. SMAC is a benchmark suite specifically designed for MARL research. It is based on the real-time strategy game StarCraft II. It offers a range of tasks and scenarios that allow researchers to test their multi-agent algorithms in complex, dynamic environments. In this environment, each allied unit is controlled by an RL agent, which can observe the distances, relative positions, unit types, and health of all ally and enemy units within its field of view at each time step. Built-in rules of the environment control all enemy units. To address the shortcomings of SMAC, SMACv2 introduces three changes: random team compositions, random starting positions, and realistic field of view and attack range. These changes encourage agents to focus more on understanding the observation space and prevent the learning of successful open-loop strategies (strategies conditioned only on the time step). Besides these changes, SMACv2 and SMAC are nearly identical in other aspects. The goal of the ally agents is to eliminate all enemy agents within a certain timeframe, and rewards are only given when enemy units are eliminated and victory is achieved. Moreover, both environments feature discrete joint action and state spaces, making them suitable for VFF methods. For more detailed settings regarding the environment parameters, please refer to Appendix~\ref{B.1}. \par

\textbf{Baseline Selection:} \cite{hu2021rethinking} achieves higher win rates by performing code-level optimizations on QMIX\cite{rashid2020monotonic} and VDN\cite{sunehag2017value}, and releases these improvements in pymarl2\footnote{https://github.com/hijkzzz/pymarl2}. Due to their excellent performance among contemporary algorithms, we include them as one of the baselines, referred to as FT-QMIX and FT-VDN. Furthermore, to address the curse of dimensionality resulting from the increased number of agents, \cite{jianye2022boosting} introduces the HPN series methods incorporating PI and PE, achieving a $100\%$ win rate in nearly all hard and super-hard SMAC scenarios. Consequently, we chose HPN-QMIX and HPN-VDN as state-of-the-art (SOTA) algorithms. To ensure fairness, all algorithms in this study are developed and tested using their open-source project, pymarl3\footnote{https://github.com/tjuHaoXiaotian/pymarl3}. Therefore, the subsequent sections will present the experimental results of six algorithms: QTypeMix, QTypeMix-B, HPN-QMIX, HPN-VDN, FT-QMIX, and FT-VDN. \par

\textbf{Evaluation Metrics:} Since the ultimate goal of both SMAC and SMACv2 is to achieve victory in battles, we use the test win rate across different scenarios as the evaluation metric. The following sections will show how the test win rate changes for different algorithms as the number of training steps increases. Each model is trained for 10,050,000 steps, with a test of 32 episodes conducted every 10,000 steps. To ensure fairness, we use the same training parameters for each algorithm in the same scenarios and do not perform detailed tuning for any specific algorithm. For detailed training parameter settings, please refer to the Appendix~\ref{B.2}. Furthermore, we conducted longer testing periods to obtain more objective and accurate results. Each trained model is tested for win rates over 1280 (32*40) episodes in the SAMC scenario and 5120 (32*160) in the SMACv2 scenario. The difference in the number of test episodes is due to the insufficient randomness in the SMAC scenario when tested in a multi-threaded parallel manner, which results in longer testing times. \par

\subsection{Experiments on SMAC}
\label{5.1}
\begin{figure*}[htbp]
	\centering
		\includegraphics[scale=0.7]{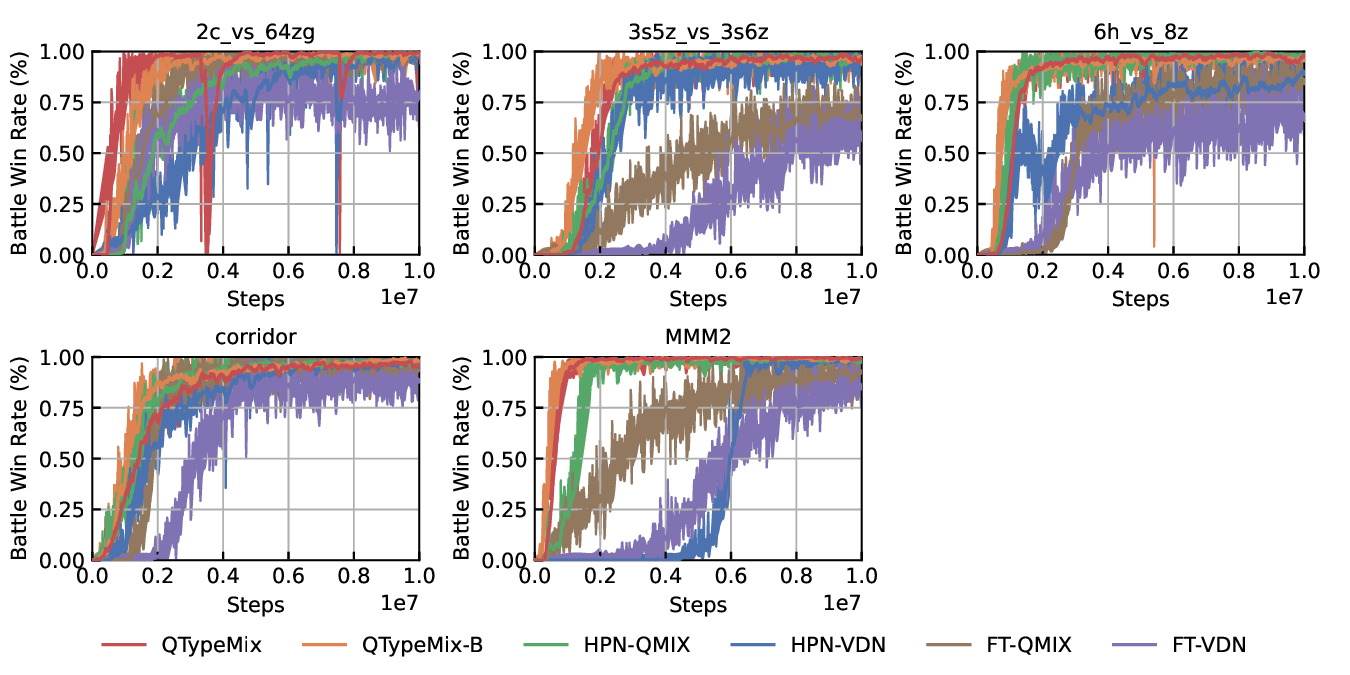}
	  \caption{Test battle win rate curves of the six algorithms during the training process on 4 SMAC maps.}\label{fig5}
\end{figure*}

Since the proposed QTypeMix is a type-based VFF method, it is foreseeable that there will be no performance improvement in scenarios involving a small number of types. In SMAC, most scenarios involve only 1 to 2 types of agents and have limited randomness. Therefore, we selected 1 hard map and 4 super hard maps for our experiments. \par

\begin{figure*}[htbp]
	\centering
		\includegraphics[scale=0.7]{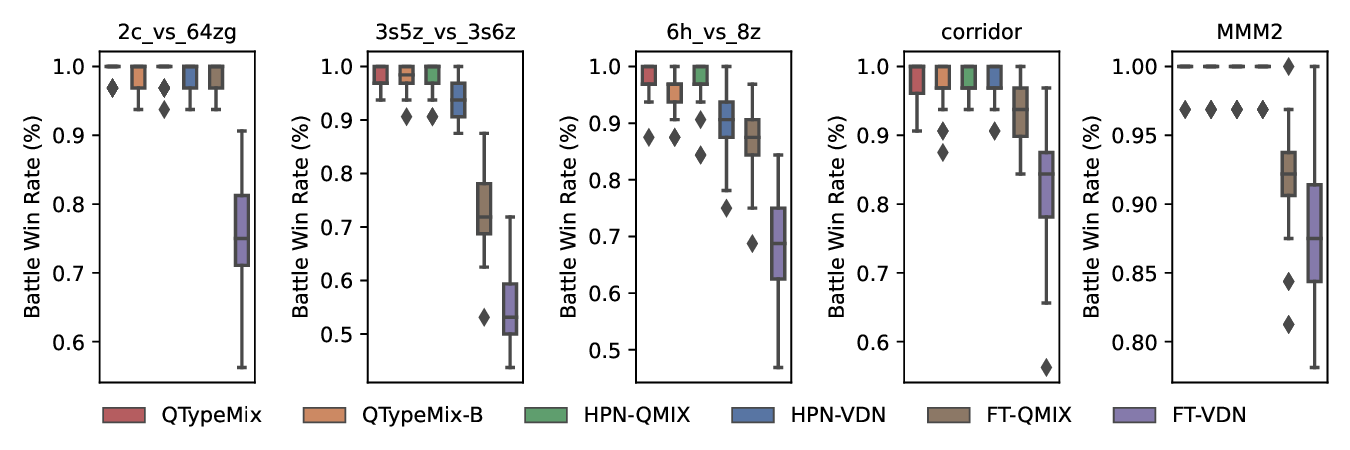}
	  \caption{Battle win rate results of 6 algorithms tested over 40 runs of 32 episodes each on 5 SMAC maps.}\label{fig6}
\end{figure*}

Fig.~\ref{fig5} presents the experimental results of six algorithms—QTypeMix, QTypeMix-B, HPN-QMIX, HPN-VDN, FT-QMIX, and FT-VDN—on 5 maps, which generally align with our expectations. Due to the relatively low difficulty of the SMAC maps, our algorithm and the SOTA algorithms almost all achieve near 100$\%$ win rates. However, differences in convergence speed can still be observed. Firstly, on maps with only one type of agent (2c vs. 64zg, 6h vs. 8z, corridor), QTypeMix shows a significant improvement in convergence rate only on 2c vs. 64zg. On the map with two types of agents (3s5z vs. 3s6z), it slightly improves the convergence rate. On the map with three types of agents (MMM2), where the convergence speed of HPN-QMIX is already breakneck at approximately 3,000,000 steps, QTypeMix and QTypeMix-B achieve convergence in around 1,600,000 steps. This indicates that QTypeMix enables the strategy to focus on practical information more quickly and efficiently, simplifying the training process. The test results of the trained models over 1280 episodes are shown in Table~\ref{tbl1}. Fig.~\ref{fig6} is a box plot drawn based on these results. QTypeMix, QTypeMix-B, and HPN-QMIX achieve SOTA performance across all tasks, with QTypeMix performing slightly better and QTypeMix-B lacking stability. HPN-VDN also reaches SOTA performance in some scenarios. In contrast, FT-QMIX and FT-VDN generally perform poorly in most cases. Overall, on SMAC maps, QTypeMix undoubtedly achieves the best convergence speed and win rate performance compared to other algorithms. However, due to the limitations of the maps, the improvement over HPN-QMIX is relatively tiny. \par

\subsection{Experiments on SMACv2}
\label{5.2}
\begin{figure*}[htbp]
	\centering
		\includegraphics[scale=0.7]{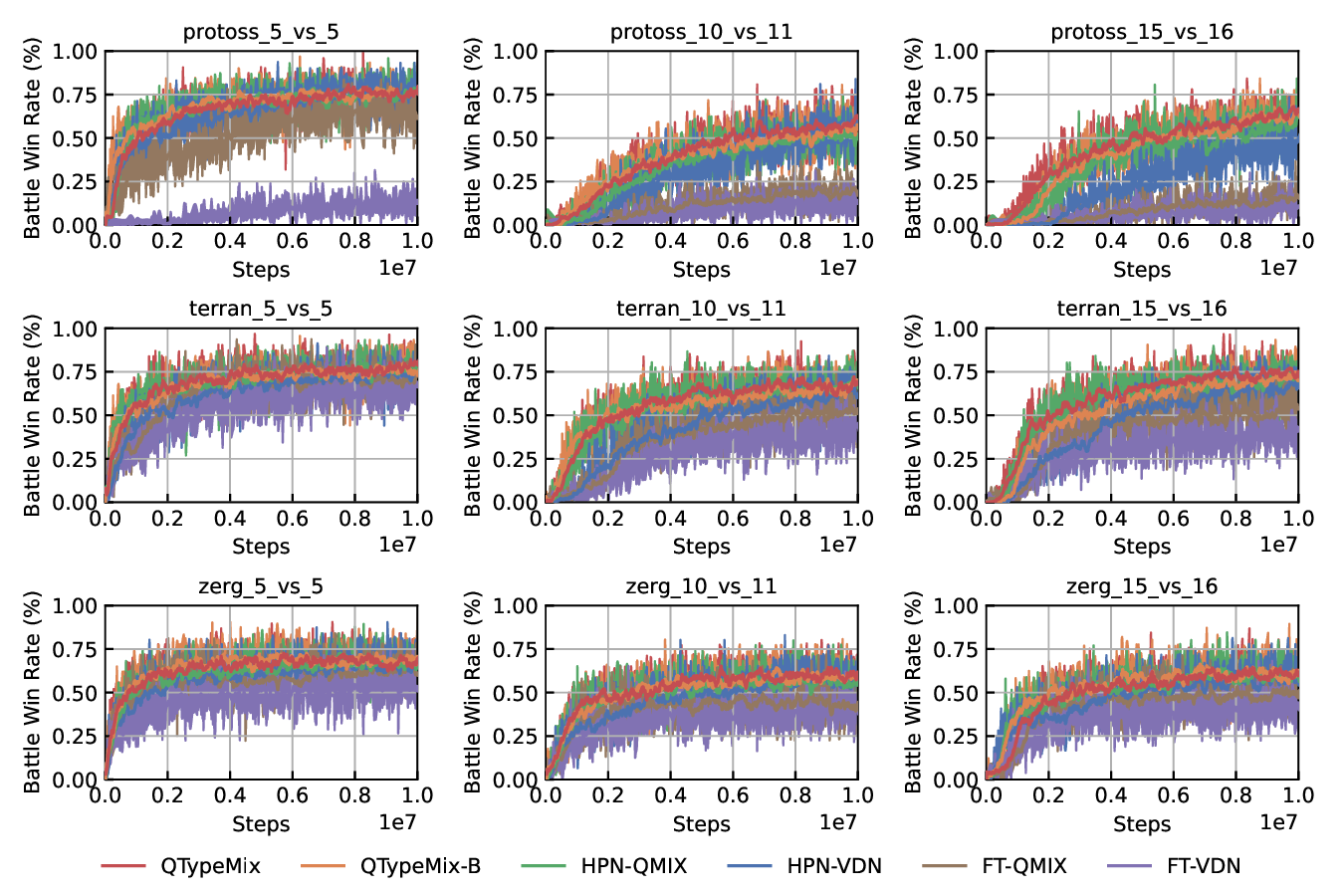}
	  \caption{Comparison of QTypeMix and QTypeMix-B with Baseline Algorithms on SMACv2.}\label{fig7}
\end{figure*}

As noted earlier, SMAC maps are not fully capable of showcasing the superior performance of QTypeMix. Therefore, we conduct experiments on 9 maps selected from SMACv2. We group these maps according to the number of ally and enemy agents: 5 vs. 5 (protoss 5 vs. 5, terran 5 vs. 5, zerg 5 vs. 5), 10 vs. 11 (protoss 10 vs. 11, terran 10 vs. 11, zerg 10 vs. 11), and 15 vs. 16 (protoss 15 vs. 16, terran 15 vs. 16, zerg 15 vs. 16). On these maps, regardless of the number of allied agents, three types of units are generated with probabilities of $[$0.45, 0.45, 0.1$]$. The significant amount of randomness introduced makes SMACv2 maps much more challenging than those in SMAC. Achieving victory is almost impossible when there is a significant disparity in the initial unit strengths or unfavorable positioning. \par

As shown in Table~\ref{tbl1} and Fig.~\ref{fig7}, the experimental results on 9 SMACv2 maps indicate that QTypeMix exhibits the highest win rate among the 6 algorithms on all maps. This is particularly evident in scenarios with a larger number of ally agents (10 vs. 11 and 15 vs. 16), where the win rate of QTypeMix significantly improves. This confirms our earlier hypothesis that QTypeMix demonstrates its superiority in scenarios with a higher number of ally agent types (although the 5 vs. 5 series maps also feature three types of agents randomly, due to the limited number of agents, often only two types are present simultaneously). On the 5 vs. 5 series maps, the performance of QTypeMix, QTypeMix-B, and HPN-QMIX is similar. The test win rates in 10 vs. 11 and 15 vs. 16 series maps follow the general trend of QTypeMix > QTypeMix-B > HPN-QMIX. Comparing the results of QTypeMix and QTypeMix-B reveals the effectiveness of extracting type embeddings. QTypeMix provides a more precise direction for the model's learning, resulting in faster convergence speeds on most maps, even though its neural network is larger. \par

It is important to note that our objective is not to achieve the highest possible win rate for any specific algorithm through meticulous hyperparameter tuning. Instead, we adopt the algorithm parameter settings from \cite{hu2021rethinking}, allowing for a fair comparison by keeping common parameters consistent across all algorithms. This approach ensures that the conclusions drawn are more objective. Although fine-tuning the training and model parameters might yield better results on SMACv2 maps, that is beyond the scope of our study. For more detailed information on model parameters, please refer to the Appendix~\ref{B.3}.

\begin{figure*}[htbp]
	\centering
		\includegraphics[scale=0.7]{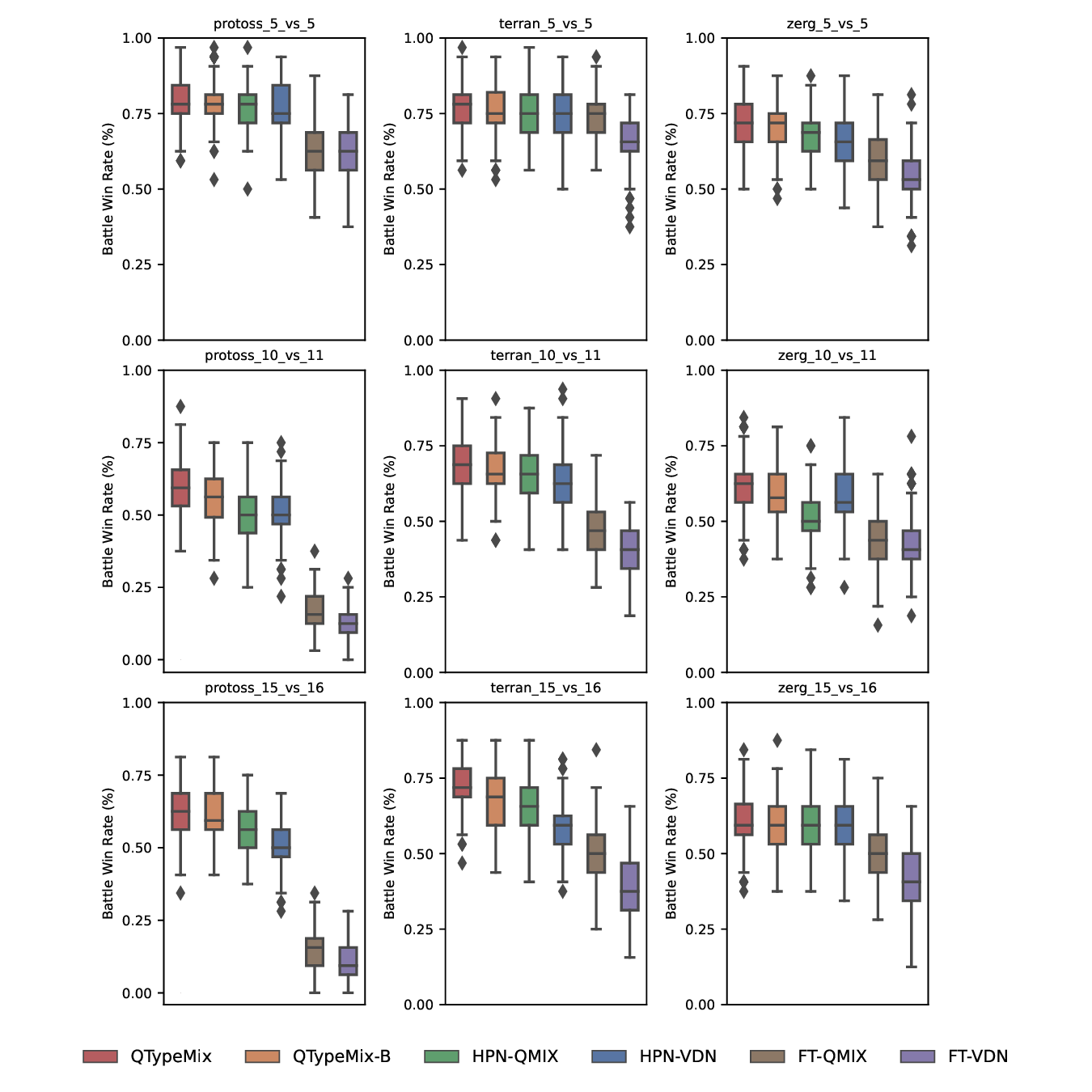}
	  \caption{Battle win rate results of 6 algorithms tested over 160 runs of 32 episodes each on 9 SMACv2 maps.}\label{fig8}
\end{figure*}

\section{Conclusion}
\label{6}

Our core concept is to enable each agent to recognize the roles they are best suited for in heterogeneous multi-agent cooperative tasks. Therefore, this paper proposes a dual-layer VFF method, QTypeMix, which introduces type information. This method divides the value decomposition process into homogeneous and heterogeneous stages based on the type-related information provided by the environment. Utilizing fine-grained value decomposition accelerates learning efficiency. Additionally, it extracts hidden features of different types from the agents' historical observations, providing a more precise direction for the training process. Experimental results on SMAC and SMACv2 indicate that QTypeMix performs well in scenarios with fewer agents and types and demonstrates impressive improvements in more complex scenarios with more agents and types. \par

However, QTypeMix also has some issues. The larger size mixers slow down the training speed of the model. In future work, we will explore filtering more practical information to reduce the model's dimensionality. Additionally, QTypeMix is just an instance of our core idea applied to the VFF method. One of our future research focuses will be on how to apply this concept to policy-based methods. \par

\section*{Acknowledgements}
This work is partially supported by the Goal-Oriented Project Independently Deployed by the Institute of Acoustics, Chinese Academy of Sciences(MBDX202106). \par

\appendix

\section{Architecture details}\label{A}

\begin{breakablealgorithm}
\caption{QTypeMix}
\label{algo1}
\begin{algorithmic}[1] 
\algsetup{linenosize=\normalsize}
\STATE Input training and model parameters; \\
\STATE Initialise $\theta = (\theta_{te}, \theta_{hpn}, \theta_{m1}, \theta_{m2})$, the parameters of mixing networks, agent networks, and hypernetworks; \\
\STATE Set the learning rate $\alpha$ and replay buffer $D = \{\}$; \\
\STATE step = 0, $\theta^- = \theta$\; \\
\WHILE{step $<$ max\_step}
    \STATE $t = 0$, $s_0$ = initial state; \\
    \WHILE{$terminal = $~False and $t < t_{max}$}
        \FOR{each agent $a$ do}
            \STATE $\tau_t^a = \tau_{t-1}^a \cup \{(o_t, u_{t-1})\}$
            \STATE $\epsilon = $ epsilon-schedule(step)
            \STATE $u_t^a = 
            \begin{cases}
                \arg\max_{u_t^a} Q(\tau_t^a, u_t^a) & \text{with probability } 1 - \epsilon \\
                \text{randint}(1, |\mathcal{U}|) & \text{with probability } \epsilon
            \end{cases}$
        \ENDFOR
        \STATE Get reward $r_t$ and next state $s_{t+1}$
        \STATE $\mathcal{D} = \mathcal{D} \cup \{(s_t, u_t, r_t, s_{t+1})\}$
        \STATE $t = t + 1$, step = step + 1
    \ENDWHILE
    \IF{$|\mathcal{D}| >$ batch\_size}
        \STATE $b \gets$ random batch of episodes from $\mathcal{D}$
        \FOR{each timestep $t$ in each episode in batch $b$ do}
            \STATE $Q_{tot} = \text{Total-Mixer}(\text{Type-Mixer}_1(Q_1(\tau_t^1, u_t^1),\ldots , Q_n(\tau_t^n, u_t^n)),\ldots,$ \\
            \STATE $\quad\quad\quad\quad\quad\quad\quad~~~~  \text{Type-Mixer}_m(Q_1(\tau_t^1, u_t^1), \ldots, Q_n(\tau_t^n, u_t^n)); \theta)$
            \STATE Calculate target $Q_{tot}$ using all Mixing-networks and the Type Embedding Extractor.
        \ENDFOR
        \STATE $\Delta Q_{tot} = r_t + \gamma \max_{u'} Q_{tot}(s',u'; \theta^-) - Q_{tot}(s,u; \theta)$ \\
        \STATE $\Delta T = ( \sum_i^n \sum_j^n I(i, j) \cdot \cos \left( E_i(Z^i, S; \theta_{te}), E_j(Z^j, S; \theta_{te}) \right)$ , $I(i, j) = 
            \begin{cases}
                -1, & \text{if } a_i \in \mathcal{T}_k, a_j \in \mathcal{T}_k \\
                1, & \text{if } a_i \in \mathcal{T}_k, a_j \notin \mathcal{T}_k
            \end{cases}$
        \STATE $\Delta \theta = \nabla_\theta (\Delta Q_{tot})^2 + \alpha \cdot \Delta T$ \\
        \STATE $\theta = \theta - \text{lr} \cdot \Delta \theta$ \\
    \ENDIF
    \IF{target\_update\_interval steps have passed}
        \STATE $\theta^- = \theta$
    \ENDIF
\ENDWHILE
\end{algorithmic}
\end{breakablealgorithm}

\section{Experiment details}\label{B}

\subsection{Environment parameters}\label{B.1}

\renewcommand{\thetable}{B.1}
\begin{table}[htbp]
\centering
\caption{Key environment, model, and training parameters.}\label{tbl2}
\begin{tabular}{
    >{\raggedright\arraybackslash}p{0.15\textwidth} 
    >{\raggedleft\arraybackslash}p{0.13\textwidth} 
    >{\raggedright\arraybackslash}p{0.15\textwidth} 
    >{\raggedleft\arraybackslash}p{0.13\textwidth} 
    >{\raggedright\arraybackslash}p{0.15\textwidth} 
    >{\raggedleft\arraybackslash}p{0.13\textwidth}
}
    \toprule
    \multicolumn{2}{c}{\textbf{Environment}} & \multicolumn{2}{c}{\textbf{Model}} & \multicolumn{2}{c}{\textbf{Training}} \\
    \cmidrule(lr){1-2}\cmidrule(lr){3-4}\cmidrule(lr){5-6}
    Parameters & Value & Parameters & Value & Parameters & Value \\
    \midrule
    difficulty & 7 & mixing\_embed\_dim & 32 & lr & 1e-3 \\
    obs\_last\_action & False & hypernet\_embed\_dim & 64 & gamma & 0.99 \\
    state\_last\_action & True & hpn\_hyper\_dim & 64 & buffer\_size & 5000 \\
    state\_timestep\_number & False & rnn\_hidden\_dim & 64 & test\_interval & 10000 \\
    conic\_fov & False & hpn\_hyper\_activation & RELU & test\_nepisode & 32 \\
    & & other\_activation & ELU & target\_update\_interval & 200 \\
    & & & & $\epsilon_\text{start}$ & 1.0 \\
    & & & & $\epsilon_\text{finish}$ & 0.05 \\
    & & & & epsilon\_anneal\_time & 100000 \\
    & & & & optimizer & adam \\
    & & & & $\alpha_{opt}$ & 0.99 \\
    & & & & max\_step & 1.005e7 \\
    \bottomrule
\end{tabular}
\end{table}

The SMAC and SMACv2 officials provide environment parameters for users to configure maps. For the sake of fairness, we primarily use the default parameters. It is important to note that since our algorithm does not target the exploration problem of agents, we set the parameter conic\_fov to False on SMACv2 maps. This means that in our experiments, agents have a circular field of view on SMACv2 maps, similar to SMAC maps. \par

\subsection{Training parameters}\label{B.2}

Training parameters in Table~\ref{tbl2} are consistent across all maps and algorithms. $\epsilon$ represents the probability of an agent taking random actions during training. Typically, a larger $\epsilon$ is used at the beginning of the algorithm to increase the diversity of samples. As we have emphasized multiple times, we do not aim to achieve better performance through fine-tuning parameters for any specific algorithm. Therefore, we adopt the parameter configurations from \cite{jianye2022boosting}, which results in some maps using unique configurations. For example, batch\_size\_run controls the number of parallel environments, set to 4 on 3s5z vs. 3s6z and 8 on other maps. td\_lambda is set to 0.3 on 6h vs. 8z and 0.6 on other maps. hpn\_head\_num is set to 2 on 3s5z vs. 3s6z and 6h vs. 8z, and 1 on other maps. Notably, hpn\_head\_num does not affect FT-QMIX and FT-VDN. \par

\subsection{Model parameters}\label{B.3}

Model parameters are used to configure the neural networks involved in the algorithm. Table~\ref{tbl2} lists some important model parameters applied to all mentioned algorithms (if needed). mixing\_embed\_dim, hypernet\_embed\_dim, hpn\_hyper\_dim, and rnn\_hidden\_dim represent the output dimension of the linear layers in the Mixers, the internal dimension of the hypernetworks in the Mixers, the internal dimension of the hypernetworks in HPN, and the hidden state dimension of the GRUs, respectively. n\_heads sets the number of attention heads for QTypeMix and QTypeMix-B. \par

\printcredits

\bibliographystyle{cas-model2-names}

\bibliography{mybib}

\end{document}